\documentclass[aps,prb,twocolumn,groupedaddress]{revtex4}
\usepackage{amsmath}
\usepackage{amssymb}
\usepackage{graphicx}

\begin{document}

\title{Finite Conductivity Minimum in Bilayer Graphene without Charge Inhomogeneities}

\author{Maxim Trushin$^{1,2}$, Janik Kailasvuori$^3$, John Schliemann$^2$, and A.H. MacDonald$^1$}

\address{$^1$Physics Department, University of Texas, 1 University Station
  C1600, Austin, 78712 Texas, USA}
\address{$^2$Institut f\"ur Theoretische Physik, Universit\"at Regensburg,
93040 Regensburg, Germany}
\address{$^3$Max-Planck-Institut f\"ur Physik komplexer Systeme,
N\"othnitzer str. 38, 01189 Dresden, Germany}

\begin{abstract}
Boltzmann transport theory fails near the linear
band-crossing of single-layer graphene and near the 
quadratic band-crossing of bilayer graphene.  
We report on a numerical study which assesses 
the role of inter-band coherence in transport when the 
Fermi level lies near the band-crossing energy of bilayer graphene. 
We find that interband coherence enhances
conduction, and that it plays an essential role in bilayer graphene's 
minimum conductivity phenomena.
This behavior is qualitatively captured by an approximate theory 
which treats inter-band coherence in a relaxation-time 
approximation.  On the basis of this short-range-disorder model 
study, we conclude that electron-hole puddle formation
is not a necessary condition for finite conductivity in bilayer graphene
at zero average carrier density. 
\end{abstract}

\maketitle

\section{Introduction}

The robust conductivity of nearly neutral graphene sheets\cite{NatMat_geim,RMP_castroneto}
is interesting from a theoretical point of view, awkward\cite{Science_geim} for some potential applications,
and among the most unexpected of graphene transport study discoveries.  As a function of 
ambipolar carrier density the minimum conductivity is $\sim e^2/h$, with relatively small sample to sample variation. 
The generally accepted explanation\cite{PNAS_adam,Cheianov_PRL,PRB_adam,PRB_rossi,Rossi_PRL,Fogler_PRL,PRB_fogler,PRB_das_sarma,PRB2_adam} for this property
starts by recognizing the influence of randomly distributed charged-impurities\cite{Nomura_PRL,Ando_JPSJ}
which induce electron-hole puddles\cite{NatPhys_martin,NatPhys_zhang}
in graphene when the global average carrier density is low.
Partly because of\cite{Cheianov_PRL} the role of 
Klein tunneling in Dirac-like systems, a network of  
conducting puddles can account for global conduction
that remains finite when the average carrier concentration falls to zero.
There are, however, indications that this explanation is
incomplete. In particular, suspended graphene\cite{NatNano_du,solidst_bolotin,PRL_bolotin}
samples still exhibit a minimum conductivity even though 
charged impurities appear to be removed upon annealing and puddle formation should
therefore be suppressed.
The present work is motivated by the view that graphene's minimum conductivity 
phenomena are more general than sometimes thought, and not necessarily associated with smooth inhomogeneities.

Independent of disorder character, transport near the band-crossing energies of graphene systems 
differs from transport near typical semiconductor band extrema in three important ways:
i) the absence of an energy gap between the conduction and valence 
bands, ii) the peculiar momentum-dependence of inter-sublattice 
hopping in graphene systems that leads to the Dirac-like electronic structure and 
iii) in the case of single-layer graphene the linear band dispersion which 
causes the two-dimensional density-of-states to vanish in the absence of disorder.  
The goal of this paper is to shed light on which of these aspects is 
responsible for conductivity minimum phenomena.  Since experiment 
indicates that there is no essential difference between the minimum
conductivity behavior of single and bilayer cases, the
dispersion law does not appear to play an essential role.
The minimum conductivity is also finite in
suspended {\em bilayer} graphene\cite{NatPhys_feldman}
samples, even though the charge carriers in this case exhibit the same parabolic\cite{PRL_mccann} dispersion that is
found in conventional two-dimensional electron systems.
We therefore focus on bilayers, and on the role of momentum-sublattice 
coupling in the absence of an energy gap.
This problem has received relatively little theoretical attention\cite{PRL_nilsson,PRB_koshino,EPJB_katsnelson,PRB_snyman,PRL_cserty,PRL_gorbachev,PRL_kechedzhi,PRB_moghaddam,PRB_culcer}.

\begin{figure}
	\centering\includegraphics[width=\columnwidth]{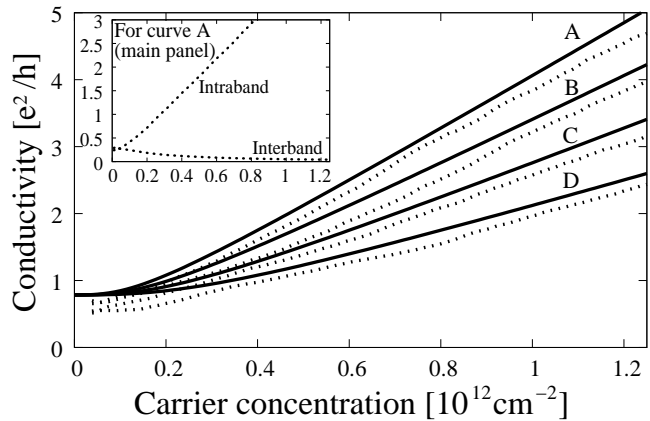}
\caption{The dotted curves depict the electrical conductivity of bilayer graphene
(per spin/valley) as a function of carrier concentration
computed according to the Kubo formula (\ref{Kubo})
for the series of model parameters specified in Table \ref{tab1}.
The solid lines correspond to the analytical approximation which is
the sum of the Drude conductivity $\sigma_\mathrm{D}$
and an interband coherent correction $\Delta\sigma$ given by Eq.~(\ref{correct}).
The inset illustrates the decomposition of the conductivity
for disorder model A into
intra- and interband coherent contributions proportional respectively to
the intra- and interband terms in the velocity operator in Eq~(\ref{Kubo}).}
	\label{fig1}
\end{figure}

Momentum-sublattice coupling in bilayers is well described by 
the $\pi$-band envelope-function effective band Hamiltonian\cite{NatMat_geim,PRL_mccann}
\begin{equation}
\label{h0}
H_0= - \frac{\hbar^2}{2m}
\left(\begin{array}{cc}
0 
& (k_x-i k_y)^2 \\ 
(k_x+i k_y)^2 & 0
\end{array} \right).
\end{equation}
Here $m \simeq 0.05 m_0$ is the effective mass, $m_0$ is the bare 
electron mass, ${\mathbf k}$ is 
the two-component particle momentum, and
the matrix structure originates from the layer and sublattice 
degrees of freedom.
The Hamiltonian $H_0$ does not contain the trigonal warping term
and just represents
the minimal model where the conductivity minimum does not vanish.
The spectrum of $H_0$ consists of parabolic conduction and 
valence bands that touch at eigenenergy $E=0$.
The sublattice degree-of-freedom is frequently viewed as a pseudospin
in order to exploit analogies between spin-orbit and pseudospin-orbit coupling.
From this point of view $H_0$ can be considered as expressing 
an effective Zeeman coupling to pseudospins that has a
strength $\hbar \Omega_k=\hbar^2 k^2/m$ which is momentum-magnitude dependent, and a
$\hat{x}$-$\hat{y}$ plane orientation angle $\phi = 2 \phi_{\mathbf k}$ where
$\phi_{\mathbf k}$ is the two-dimensional momentum direction.
The pseudospin precession axis therefore changes whenever an electron 
is scattered between momentum states. 
When the precession frequency $\Omega_k$ is larger
than the momentum scattering rate $\tau^{-1}$,
the pseudospin precesses a few times between
collisions and any initial transverse component is likely to 
be randomized.  The conductance minima phenomena occur for 
energies $E$ near zero for which $\Omega_k \tau$ is always small
and pseudospin components transverse to the precession axis 
are not expected to randomize.  This observation alone suggests 
the possibility that atypical quantum effects could play a role.
This is what we can see in Fig.~\ref{fig1}:
The conductivity never falls to zero
for any reasonable choice of parameters as long as the interband coherence
is included in the model, even when charged impurities are absent
and electron-hole puddle formation\cite{PRB_adam,PRB_das_sarma} is not expected.
We focus solely on the zero temperature limit.
The finite temperature\cite{PRB_adam,PRB2_adam,PRB_lv} can lead to the thermally excited
carriers which may spoil the interband coherence effect.
The intervalley scattering is also assumed to be absent here.

\section{Kubo and Boltzmann Theories}

We have evaluated the conductivity numerically 
using the non-interacting particle Kubo formula expression.
This approach has the advantage that it is exact\cite{PRL_nomura},
or at least would be if computational resources were infinite.  On the other hand 
it does not lend itself to a satisfying qualitative understanding.
We therefore compare our numerical results with those 
predicted by a heuristic semiclassical theory\cite{PRL_trushin,PRB_auslender}
that captures inter-band coherence corrections to the Boltzmann equation.
We first comment briefly on these two approaches.

The finite-size Kubo formula for the static conductivity is,
\begin{equation}
\label{Kubo}
\sigma_\mathrm{K}=-\frac{i\hbar e^2}{L^2}\sum\limits_{n,n'}
\frac{f^0_{E_n}-f^0_{E_{n'}}}{E_n - E_{n'}}\frac{\langle n \vert v_x \vert n'\rangle\langle n' \vert v_x \vert n\rangle}{E_n - E_{n'} + i\eta},
\end{equation}
where $\mathbf{v}$ is the velocity operator, $f^0_{E_n}$ is the
Fermi-Dirac distribution function, and $\vert n\rangle$ denotes an exact eigenstate
of the Schr\"odinger equation for a finite-size disordered system with 
periodic boundary conditions: $(H_0+U)\psi_n= E_n \psi_n$
with $U(\mathbf{r})=u_0\sum_i^{N_s} \delta(\mathbf{r}-\mathbf{R}_i)$
for the short-range disorder model we consider.
The scattering locations $\mathbf{R}_i$ and potential signs
are random. We solve the Schr\"odinger equation using a 
large momentum-space cutoff $k^* \approx \sqrt{5\cdot 10^{13}}\,\mathrm{cm}^{-1}$
which corresponds to the energy scale at which the split-off bands of bilayer 
graphene become relevant and our two-band model no-longer applies.

The physical {\em dc} conductivity can be obtained from Eq.~(\ref{Kubo})
by extracting the limit in which the system size first approaches $\infty$,
and then $\eta$ approaches zero maintaining a value larger than the 
typical level spacing $\delta E$.  For the model considered here 
$\delta E=(2\pi\hbar^2)/(mL^2)$ where $L^2$ is the finite-size system area.
The finite value of $\eta$ can be understood as representing energy uncertainty 
due to the finite lifetime of electrons in a system coupled to source and 
drain reservoirs.  To eliminate the influence of the bath on
the conductivity itself, the momentum relaxation time $\tau$
due to internal scatterers must be much smaller
than $\hbar/ \eta$ \cite{JPC_licciardello}.
We estimate $\tau$ using the Fermi golden-rule expression:
$\tau=2\hbar^3/m n_s u_0^2$ where $n_s=N_s/L^2$ is the impurity density.
Since the smallest possible $\delta E$ is limited by 
numerical practicalities, we can estimate the conductivity only for 
relatively strong disorder.
Conductivities obtained directly from Eq.~(\ref{Kubo})
undergo the phase coherent fluctuations; we simulate macroscopic system
conductivities by averaging the conductivity over an energy interval
containing $10$--$100$ levels, over boundary conditions, 
and over several disorder potential realizations.
Note, that the conductivity fluctuation amplitude
turns out to be essentially smaller that $e^2/h$ near the neutrality
point. This makes our numerical approach reliable
for the conductivity minimum evaluation.

\begin{table}
\begin{tabular}{|c|c|c|c|c|c|}
\hline
Label &   $\tau$    &  $\mu$    & $n_s$   &  $\eta\tau/\hbar$ \\
&($10^{-13}\,\mathrm{s}$)& ($10^{3}\,\mathrm{cm}^{2}/\mathrm{Vs}$) & ($10^{12}\,\mathrm{cm}^{-2}$) & at $\eta=10\delta E$ \\ \hline
A     & 0.30  &  1     & 0.81  & 0.13 \\
B     & 0.25  &  0.83  & 0.97  & 0.10 \\
C     & 0.20  &  0.66  & 1.22  & 0.08 \\
D     & 0.15  &  0.50  & 1.62  & 0.06 \\ \hline
\end{tabular}
\caption{Parameters for Fig.~\ref{fig1}: 
$\tau$ is the momentum relaxation time,
$\mu=e\tau/m$ is the mobility of carriers,
$n_s$ is the concentration of 
short range scatterers with strength fixed at a value  
$u_0=\pi^2 \hbar^2/5m$
small enough to validate the golden-rule life-time expression, and
$\delta E=2\pi\hbar^2/L^2 m$ is the level spacing
for sample size $L=1.8\times 10^{-5}\,\mathrm{cm}$.
At this sample size dependence on $L$ is weak.
The momentum cut-off $k^*$ and $L$ fix the
Hamiltonian matrix dimension at $3362\times 3362$.
The computations have been performed at zero temperature.}
\label{tab1}
\end{table}

Below we compare our numerical results for the conductivity to an analytic
modified Boltzmann equation theory.
When coherence effects are retained the distribution function $f({\mathbf k})$
becomes a $2\times 2$ matrix with band labels.
The steady state limit of its equation of motion is 
\begin{equation}
\label{keq}
\frac{1}{\hbar}\left\{e{\mathbf E} \frac{\partial f({\mathbf k})}{\partial{\mathbf k}} + i\left[H_0,f({\mathbf k})\right]
\right\}
=\mathrm{I}[f({\mathbf k})],
\end{equation}
where ${\mathbf E}$ is an electric field small enough to justify
linear-response theory,
$\mathrm{I}[f({\mathbf k})]$ is the collision 
integral which accounts for 
disorder scattering, and the commutator
$\left[H_0,f({\mathbf k})\right]$ accounts for the difference in 
time evolution between conduction and valence band eigenstates.
When the collision integral is evaluated to leading 
(second) order in the (configuration averaged) impurity potential, the collision term
(including its off-shell terms\cite{PRB_auslender,arxiv_lueffe})
reduces to the simple matrix relation-time form, $\mathrm{I}[f({\mathbf k})] \to 
- f^{(1)}({\mathbf k})/\tau$, where $f^{(1)}$ is the deviation from equilibrium.
This is a remarkable property of the two-band bilayer model with 
$\delta$-function scatterers.
In the $H_0$ eigenstate basis, 
the density-matrix linear response $f^{(1)}$ then reads
\begin{equation}
\label{solut}
f^{(1)}= \tau e{\mathbf E} 
\left(\begin{array}{cc}
\mathbf{v}_{++}\left(-\frac{\partial f^0_{E_{k+}}}{\partial E_{k+}}\right) 
& \mathbf{v}_{+-}\frac{f^0_{E_{k-}}-f^0_{E_{k+}}}{\hbar\Omega_k (1+i\Omega_k\tau)} \\ 
\mathbf{v}_{-+}\frac{f^0_{E_{k-}}-f^0_{E_{k+}}}{\hbar\Omega_k (1-i\Omega_k\tau)} &
 \mathbf{v}_{--}\left(-\frac{\partial f^0_{E_{k-}}}{\partial E_{k-}}\right)
\end{array} \right).
\end{equation}
Here, $E_{k\pm}$ are the eigenvalues of $H_0$,
$\Omega_k = \hbar k^2/m$, and 
$\mathbf{v}_{\sigma,\sigma'}$ is the velocity operator written in the
$H_0$ eigenstate basis.
Given this approximation for the linear response of the distribution function,
it is easy to calculate the Boltzmann conductivity: 
$\sigma_\mathrm{B}=j_x/E_x$ where $j_x$ is the electrical current,
$\mathbf{j}=e\int\frac{d^2 k}{(2\pi)^2}\mathrm{Tr}\left[\mathbf{v}f^{(1)}(\mathbf{k})\right]$.
Note that neither $\mathbf{v}$ nor $f^{(1)}(\mathbf{k})$ are diagonal, and that
$\mathbf{j}$ therefore includes interband coherence contributions.
The intraband contribution to the conductivity stems from
the diagonal terms in Eq.~(\ref{solut}) and
is given by the simple Drude formula $\sigma_\mathrm{D} = e^2 n\tau/m$,
where $n$ is the carrier concentration $n=k_F^2/(4\pi)$
with $k_F$ being the Fermi momentum.

\section{Results} 

Numerical results for the dependence of Kubo conductivity 
on carrier density are presented in Fig.~\ref{fig1} for a series of model
parameter values summarized in Table I.
Our main finding is that the conductivity remains finite as the 
carrier density approaches zero.  We do not observe any systematic
dependence of the minimum conductivity,
$\sigma_\mathrm{min} \sim 0.7 e^2/h$ per spin and valley,
on model system parameters.


There are two elements in our model which couple
the two bands in the Hamiltonian (\ref{h0}) and both are important
for the conductivity minimum phenomena.
The first is the velocity operator $\mathbf{v}_{\sigma,\sigma'}$.
The second is the scattering potential $U(\mathbf{r})$
which can produce interband scattering.
We quantify the role of interband coupling by separating
both velocity operators in Eq.~(\ref{Kubo}) into intra-band and 
inter-band contributions to express the conductivity as the sum of 
intra-band ($\propto \mathbf{v}_{\pm\pm}\mathbf{v}_{\pm\pm}$),
inter-band  ($\propto \mathbf{v}_{\pm\mp}\mathbf{v}_{\mp\pm}$),
and interference ($\propto \mathbf{v}_{\pm\pm}\mathbf{v}_{\pm\mp}$) terms.
We find that the interference terms 
average to negligible values.  As illustrated in Fig.~\ref{fig1}(inset), the 
intra-band contribution dominates in the higher carrier density 
Boltzmann transport regime, as expected.
However, it does not completely vanish at zero density as long as the scattering
potential is in play. The inter-band contribution, in contrast,
increases substantially near the neutrality point.
Fig.~\ref{fig1}(inset) shows that
$\sigma_\mathrm{min}$ is due substantially, and possibly dominantly, to the non-classical 
interband coherent contribution.


\begin{figure}
	\centering

	\includegraphics[width=\columnwidth]{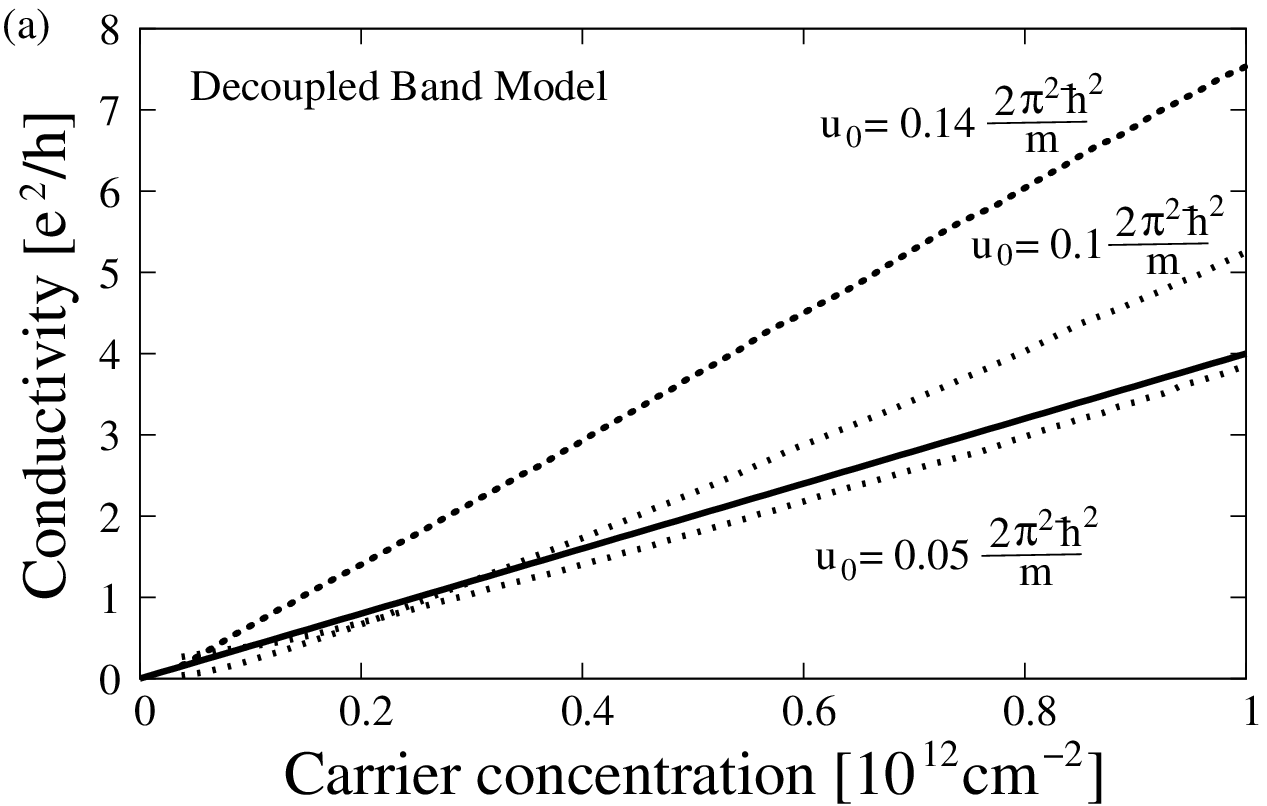}
	\includegraphics[width=\columnwidth]{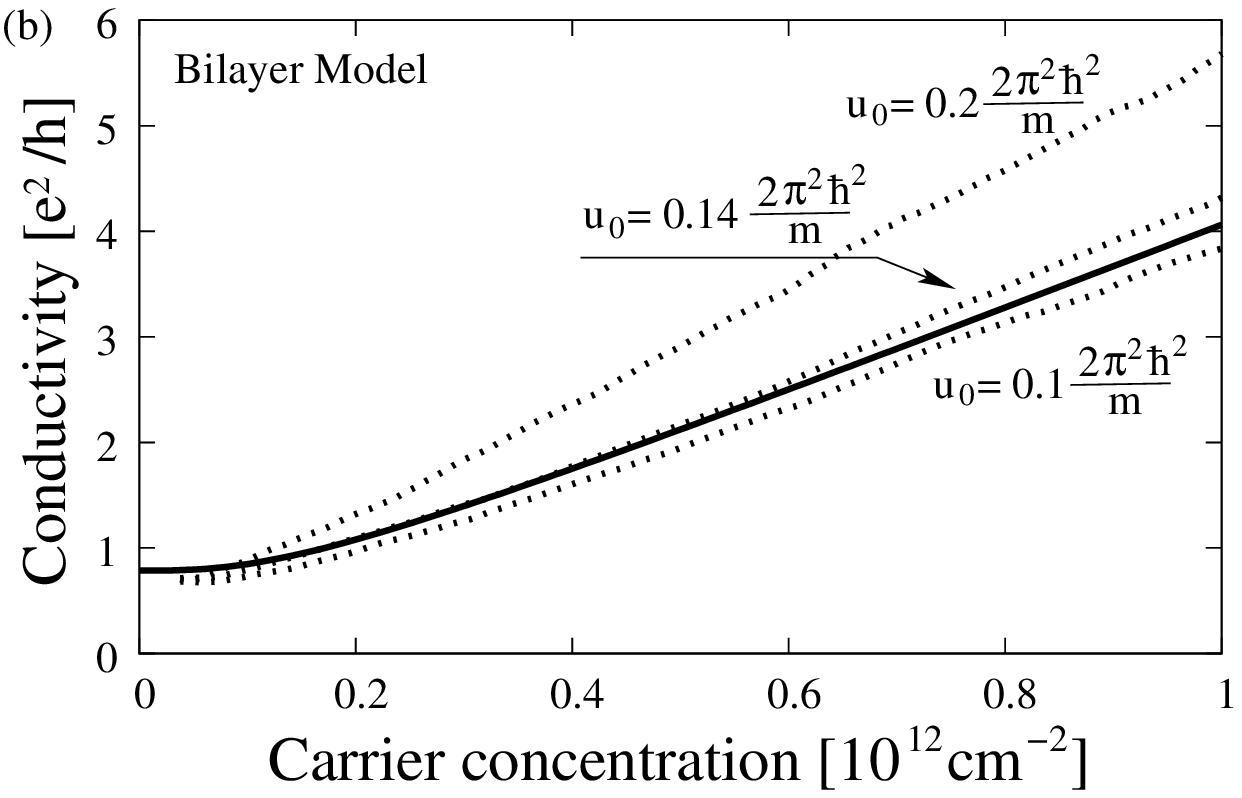}
\caption{Comparison between Kubo conductivities (\ref{Kubo}) of
(a) the decoupled band model  and (b) bilayer  graphene.
These results were obtained for a series of models with 
identical golden-rule relaxation times $\tau=0.3\cdot 10^{-13}\,\mathrm{s}$,
and sample sizes $L=1.8\times 10^{-5}\,\mathrm{cm}$.
(The concentration of scatterers $n_s$ was adjusted appropriately in each case.)
One can see that the conductivity minimum for the decoupled band model
vanishes whereas for the bilayer model it is finite and
insensitive to the scattering potential strength.
The thick solid lines show the naive prediction of (a) Drude theory
and (b) our interband coherent Boltzmann model
with golden-rule relaxation times.}
	\label{fig2}
\end{figure}

In an attempt to isolate the source of the peculiar conductivity behavior
we have in Fig.~(\ref{fig2}) compared the numerical conductivities 
of our bilayer model with those of a decoupled 
band model in which $H_0 \to \hbar^2(k_x^2+k_y^2)\sigma_z/2m$.
The two models have the same density-of-states, but the
decoupled band model has no 
interband velocity-operator matrix elements,
and the scattering potential $U(\mathbf{r})$
is not able to couple the bands.
The golden-rule relaxation times of the models are identical when we also let 
$u_0\to u_0/\sqrt{2}$
to compensate for the suppression of right-angle scattering in the bilayer case.
Fig.~\ref{fig2}a shows that $\sigma_\mathrm{min} \to 0$ in the decoupled band model.
Deviations from the Drude formula at low carrier concentrations 
in Fig.~\ref{fig2}a have a negative sign
and are consistent with Anderson insulator behavior.
In Fig.~\ref{fig2} we also see enhanced conductivity compared to the 
Boltzmann model at larger values of $u_0$ at high carrier 
densities, which we attribute simply to an overestimate of 
scattering rates by the golden-rule expression.
The small negative deviation from the Boltzmann
model at small $u_0$ may partially reflect weak localization \cite{PRL_kechedzhi,PRL_gorbachev}.

In the zero-temperature limit of the generalized Boltzmann theory,
the integrals over wavevector in the expression for the interband-coherence conductivity
can be evaluated to obtain
\begin{equation}
\label{correct}
\Delta\sigma = \frac{e^2}{2 h}\left[\frac{\pi}{2}
-\mathrm{tan^{-1}}\left(\Omega_{k_F}\tau\right)\right],
\end{equation}
and the total Boltzmann conductivity will be 
$\sigma_\mathrm{B}=\sigma_\mathrm{D}+\Delta \sigma$.
It follows that $\sigma_\mathrm{B}$ never falls down below 
$\sigma_\mathrm{min}=\pi e^2/4h$ for any choice of parameter values.
This value agrees with Ref.~\cite{PRB_culcer},
where a related modified Boltzmann approach is combined with a 
four-band effective Hamiltonian for the carriers,
as well as with recent
theoretical predictions\cite{arxiv_lueffe} using other closely related approaches.
Our $\sigma_\mathrm{min}$ differs from
the one obtained for {\em ballistic} bilayer graphene\cite{EPJB_katsnelson,PRB_snyman,PRB_cserti},
where the $\sigma_\mathrm{min}$ is attributed to evanescent modes
penetrating the sample from contacts.
We emphasize that Eq.~(\ref{correct}) should be only 
seen as the rough analytical approximation for
our numerical results. Eq.~(\ref{correct})
together with the Drude term
fits the numerical conductivity curves quite well, but it
does not mean that the conductivity minimum is {\em exactly} $\pi e^2/4h$.
However, the similarity of $\sigma_\mathrm{min}$ values
obtained with different approximate approaches
might suggest a common underlying origin
in a relationship to the spectral flows associated
with the topological properties\cite{Nomura_PRL2,PRL_bardarson,PRL_cserty}
of graphene single-layer and bilayer bands.

To conclude, in our approximate theory the
minimum conductivity is mainly due to a electric field driven  
coherence between the conduction and valence bands.
Momentum space drift due to the electric field does not repopulate  
momenta in a full valence band,
as maintained in text-book transport theory, but it does drive the  
system from equilibrium in that it alters the relationship between  
momentum and sublattice pseudospin.
There is still exactly one electron at each momentum, but  the  
momentum states no longer come with definite helicity, {\it i.e.}   
are no longer the equilibrium valence band wavefunctions.
As consequence, all valence electrons contribute to the conductivity,
although the contribution from large momenta
($\Omega_{k} \tau > 1$) gets suppressed by the larger spin precession.
Our numerical calculation provide at least partial support for this  
picture.
Quantitative discrepancies might come from not accounting for weak  
localization effects and  the influence of disorder on the  
equilibrium state.


\section{Summary}

We have used numerical exact diagonalization calculations to demonstrate 
i) that the conductivity of bilayer graphene in the limit of zero carrier-density
$\sigma_\mathrm{min} \sim e^2/h$,
ii) that inter-band coherence response plays a key role in this 
property, and 
iii) that the formation of electron-hole puddles due to strong but 
smooth potential variations is not a necessary condition for the
minimum conductivity phenomena.
We believe that our model is relevant to suspended graphene samples in 
which charged impurities are removed by annealing.
When spin and valley degeneracy is accounted for we estimate 
numerically $\sigma_\mathrm{min}^{-1} \sim 8.2\,\mathrm{k\Omega}$
which appears to be consistent with current measurements\cite{NatPhys_feldman}.

\noindent
{\em Acknowledgments} ---
This work was funded by DFG through the project TR 1019/1-1 (M.T.).
J.S. was supported by DFG via GRK 1570.
A.H.M. was supported by the Welch Foundation
(grant \#F-1473) and by the NSF-DMR program.

\bibliography{graphene.bib}

\end{document}